\documentstyle[12pt]{article}
\begin{document}
\newcommand{\de}{\delta}\newcommand{\ga}{\gamma}
\newcommand{\e}{\epsilon} \newcommand{\ot}{\otimes}
\newcommand{\be}{\begin{equation}} \newcommand{\ee}{\end{equation}}
\newcommand{\ba}{\begin{array}} \newcommand{\ea}{\end{array}}
\newcommand{\beq}{\begin{equation}}\newcommand{\eeq}{\end{equation}}
\newcommand{\tmod}{{\cal T}}\newcommand{\amod}{{\cal A}}
\newcommand{\bemod}{{\cal B}}\newcommand{\cmod}{{\cal C}}
\newcommand{\dmod}{{\cal D}}\newcommand{\hmod}{{\cal H}}
\newcommand{\s}{\scriptstyle}\newcommand{\tr}{{\rm tr}}
\newcommand{\einsop}{{\bf 1}}
\def\oR{R^*} \def\upa{\uparrow}
\def\R{\overline{R}} \def\doa{\downarrow}
\def\dag{\dagger}
\def\ve{\epsilon}
\def\si{\sigma}
\def\ga{\gamma}
\newcommand{\reff}[1]{eq.~(\ref{#1})}
\centerline{\bf{ALGEBRAIC PROPERTIES OF AN INTEGRABLE }}
\centerline{\bf{$t-J$ MODEL WITH IMPURITIES}}
~~~\\
 \begin{center}
{\large Angela Foerster$^{1}$, Jon Links$^{1,2}$ and Arlei Prestes Tonel$^{1}$}
~~~\\
{$ \phantom{0000000000}$}

{\em ${}^1$Instituto de F\'{\i}sica da UFRGS \\ Av. Bento
Gon\c{c}alves
9500,
Porto Alegre, RS - Brazil }
~~~\\
{$ \phantom{0000000000}$}

{\em ${}^2$Department of Mathematics \\
University of
Queensland, Queensland, 4072,  Australia}
\end{center}
~~~\\
\begin{abstract}
We investigate the algebraic structure of a recently proposed
integrable $t-J$ model with impurities. Three forms of the Bethe
ansatz equations are presented corresponding to the three choices for
the grading. We prove that the Bethe ansatz states are highest weight vectors
of the underlying $gl(2|1)$ supersymmetry algebra. By acting with the
$gl(2|1)$ generators we construct a complete set of states for the model.
\end{abstract}
\vfil\eject
\begin{flushleft}
Corresponding author:
{\bf Prof.  A. Foerster} \\
Address: Instituto de F\'{\i}sica da UFRGS, \\
$\phantom{00000000}$ Av.Bento Gon\c{c}alves 9500, \\
$\phantom{00000000}$ 91.501-970-Porto Alegre, RS - BRAZIL \\
e-mail:angela@if.ufrgs.br \\
Telephone: 5551 3166471 \\
Fax: 5551 3191762 \\
\vspace{1cm}
\end{flushleft}

\vfil\eject
\section*{\bf 1. Introduction}

The recent interest in the study of impurities coupled to both Fermi and
Luttinger liquids has lead to the introduction of new theoretical approaches
in order to study such systems. One approach
is to adopt the prescription of the Quantum Inverse Scattering Method (QISM)
which permits impurities to be incorporated into a model in such a way that
integrability is not violated. The origins of this method trace back to the work
of Andrei and Johannesson \cite{joh} who constucted an integrable
version of a spin 1/2
Heisenberg chain containing an arbitrary spin $S$ impurity (see also
\cite{sse,hpe}).

The long studied $t-J$ model describes strongly correlated electrons
interacting on a lattice comprising nearest neighbour hopping and spin
exchange interactions. The model is defined on a restricted
Hilbert space such that double occupancy of electrons at a single site
is forbidden.
On a one-dimensional lattice with periodic
boundary conditions there exists a special choice of $t,\,J$ such that
the model can be solved exactly \cite{lai,su,schlot}. Subsequently, it
was shown that at this point the model acquires supersymmetry with respect
to the Lie superalgebra $gl(2|1)$ \cite{sar}. The supersymmetry reflects
the fact that the model possesses both bosonic and fermionic degrees of
freedom. The existence of solvability and supersymmetry can be
understood by the fact that the model can be derived in the context of
the QISM \cite{ek,fk} which also proves integrability.
Application of the {\it algebraic} Bethe ansatz leads to
three different sets of Bethe ansatz equations corresponding to
different choices of the grading.
Furthermore, theoretical techniques based on the underlying
supersymmetry expose significant consequences
for the multiplet structure of the model. Specifically, it was shown in
\cite{fk} that the Bethe states are highest weight states of the $gl(2|1)$
superalgebra and that a complete set of states could be obtained by
exploiting this supersymmetry.

Adopting the QISM, Bed\"urftig et. al. \cite{bef} introduced an integrable
periodic $t-J$ model with impurity. In this model, the nature of the
impurity is such that doubly occupied states can occur
at those sites at which the impurities are situated.
The origin of the impurity
comes from changing the representation of the $gl(2|1)$ generators
at some lattice sites from the fundamental three-dimensional
representation to the one parameter family of typical four dimensional
representations which were introduced in \cite{bglz} to derive the
supersymmetric $U$ model. The resulting impurity model is significant in that
the parameters from the four dimensional representations can be varied
continuously without losing integrability. The impurities couple to both spin
and charge degrees of freedom and these parameters describe the strength of
the coupling of the impurities to the host chain. Limiting cases describe
effective supersymmetric $t-J$ models with additional sites or decoupled
impurities in a $t-J$ chain with twisted boundary conditions. Using the
method of the algebraic Bethe ansatz, a solution for a particular choice of
grading was derived. Also, a comprehensive
analysis of thermodynamic properties such as ground state structure,
elementary excitations, low-temperature specific heat and transport
properties was given in \cite{bef}.

In this paper, we undertake an investigation into the detailed algebraic
aspects of this impurity model. First, we present all three forms
of the algebraic Bethe ansatz solutions corresponding to the three different choices
of grading; viz. FBB, BBF, BFB. The case of FBB is that given in \cite{bef}.
A significant point here is that only
recently have new Bethe ansatz methods been proposed in order to solve
the cases of the other gradings because of a lack of a suitable (unique) reference state.
Rather one is forced to work with a subspace of reference states. This
approach was developed in the works of Abad and R{\'{\i}}os
for an $SU(3)$ impurity spin chain \cite{ar1,ar2} and has already
been adopted in \cite{pf} to find a Bethe
ansatz solution of the supersymmetric $U$ model starting from a
ferromagnetic space of states and in the works \cite{ar3,angi5,marcio} regarding
another integrable $t-J$ model with impurities. At this point
we mention that other impurity $t-J$ models have been studied.
In the work of Bares impurities were introduced into the model
by way of inhomogeneities in the transfer matrix of the
system \cite{b}. Kondo (magnetic) impurities for arbitrary spin
$S$ have been studied by Schlottmann and Zvyagin \cite{zvy}.
These works treat cases of bulk impurities with periodic boundary conditions.
It is also possible to consider open chains where the impurities
are situated on the boundary. In these cases the impurity interactions
assume very different forms than the case of bulk impurities and
examples of these models have been considered in works of
Wang et al \cite{pu} and Zhou et al \cite{hq}.

We also show that the Bethe ansatz states are highest weight states
with respect to the underlying $gl(2|1)$ supersymmetry algebra. By acting
with the $gl(2|1)$ lowering operators on
the Bethe states we obtain additional eigenvectors. Then by a combinatorial
argument we claim that in fact we recover a complete set of states for the
model. Aside from the (non-impurity) integrable $t-J$ model, completeness
of the Bethe ansatz solution has been studied for example for the cases of the
Heisenberg model \cite{hb,fad}, the Hubbard model \cite{fab1} and the $gl(2|2)$
supersymmetric extended Hubbard model \cite{schout}. Completeness for the integrable
spin $S$ impurity Heisenberg model was discussed by  Kirillov \cite{kir1}. Our
results give the first example of completeness for 
a supersymmetric, integrable impurity model.

The paper is organized as follows. In section 2 we recall the
construction of the model through the QISM and also fix notation.
In section 3 we discuss the algebraic Bethe ansatz solutions
of this model. The highest weight property of the
Bethe states is demonstrated in section 4 and the completeness
of the Bethe states is shown in section 5.

\section*{\bf 2. Quantum Inverse Scattering Method}

Recall that the Lie superalgebra $gl(2|1)$
has generators $\{E_j^i\}_{i,j=1}^{3}$ satisfying the commutation relations
\be  [E^i_j,\,E^k_l]=\delta^k_jE^i_l-(-1)^{([i]+[j])([k]+[l])}
\delta^i_lE^k_j  \label{cr}\ee
where the ${\bf Z}_2$-grading on the indices is determined by
\begin{eqnarray}
[i]&=&0~~\rm {for} ~~ i=1,2,   \nonumber \\
{[i]}&=&1~~\rm {for} ~~ i=3.   \nonumber
\end{eqnarray}
This induces a ${\bf Z}_2$-grading on the $gl(2|1)$ generators
through
$$\left[E^i_j\right]= ( [i]+[j]) \,(\rm{mod} 2). $$

The vector module $V$ has basis $\{v^i\}_{i=1}^{3}$ with action defined
by
\be
E^i_jv^k=\delta^k_jv^i. \label{vr}
\ee
Associated with this space there is a solution $R(u)\in \rm {End}
(V\otimes V)$ of the Yang-Baxter equation
\be R_{12}(u-v)R_{13}(u)R_{23}(v)=R_{23}(v)R_{13}(u)R_{12}(u-v)
\label{yb}
\ee
on the space $V\otimes V\otimes V$ which is given by
\be
R(u)=I\otimes I -\frac 2u \sum_{i,j}e^i_j\otimes e_i^j (-1)^{[j]}.
\label{rm1}
\ee
We remark that  \reff{yb} is acting on a supersymmetric space so the
multiplication of tensor products is governed by the relation
\be
(a\otimes b)(c\otimes d)=(-1)^{[b][c]}ac\otimes bd  \label{mult}
\ee
for homogeneous operators $b,c$.

The solution given by \reff{rm1} allows us to construct a universal
$L$-operator which reads
\be L(u)=I\otimes I -\frac 2{u-\lambda} \sum_{i,j}e^i_j\otimes E^j_i (-1)^{[j]}
\label{lop} \ee
for any complex parameter $\lambda$. 
This operator gives us a solution of the Yang-Baxter equation of the
form
$$ R_{12}(u-v)L_{13}(u)L_{23}(v)=L_{23}(v)L_{13}(u)R_{12}(u-v)$$
on the space $ V\otimes V\otimes gl(2|1)$ which follows from the
commutation relations \reff{cr}.
We use the one-parameter family of four-dimensional representations
of $gl(2|1)$ \cite{bglz} acting on the module $W$ with basis
$ \{v^a, v^b, v^c, v^d \}$.
The explicit matrix representatives are given by
\begin{eqnarray}
&&E^1_2=\left|b\right>\left<c\right|,\nonumber \\
&&E^2_1=\left|c\right>\left<b\right|,\nonumber \\
&&E^1_1=-\left|c\right>\left<c\right|-\left|d\right>\left<d\right|,\nonumber \\
&&E^2_2=-\left|b\right>\left<b\right|-\left|d\right>\left<d\right|,\nonumber \\
&&E^2_3=\sqrt{\alpha}\,\left|a\right>\left<b\right|+\sqrt{\alpha+1}\,\left|c\right>\left<d\right|,\nonumber \\
&&E^3_2=\sqrt{\alpha}\,\left|b\right>\left<a\right|+\sqrt{\alpha+1}\,\left|d\right>\left<c\right|,\nonumber \\
&&E^1_3=-\sqrt{\alpha}\,\left|a\right>\left<c\right|+\sqrt{\alpha+1}\,\left|b\right>\left<d\right|,\nonumber \\
&&E^3_1=-\sqrt{\alpha}\,\left|c\right>\left<a\right|+\sqrt{\alpha+1}\,\left|d\right>\left<b\right|,\nonumber \\
&&E^3_3=\alpha\,\left|a\right>\left<a\right|+(\alpha+1)\,\left (\left|b\right>\left<b\right|+\left|c\right>\left<c\right|\right )
  +(\alpha+2)\,\left|d\right>\left<d\right|.\label{matrix}
\end{eqnarray}
This representation holds for any complex value of $\alpha$.
However in the following we will assume that $\alpha$ is real and positive
ensuring that the Hamiltonian is Hermitian.
By taking this representation in the expression \reff{lop} we obtain
an $R$-matrix, denoted by $R^*(u)$, which satisfies the following
Yang-Baxter equation
\be
R_{12}(u-v)\oR_{13}(u)\oR_{23}(v)=\oR_{23}(v)\oR_{13}(u)R_{12}(u-v)
\label{yb1}
\ee
acting on $V\otimes V\otimes W.$

The impurity model is constructed with generic quantum spaces
represented by $V$ and the impurity spaces $W$.
To this end take
some index set $I=\{p_1,p_2,....,p_l\},~1\leq p_i\leq L$ and define
$$ X=\bigotimes_{i=1}^L X_i$$
where
\begin{eqnarray}
X_i&=&V~~~\,~~~~~~~\rm{if} ~~i\notin I,  \nonumber \\
X_i&=&V\otimes W~~~\rm{if} ~~i\in I. \label{w}
\end{eqnarray}
In other words for each $i\in I$  we are coupling an impurity into the lattice
which will be situated between the sites $i$ and $i+1$.

Next we define the monodromy  matrix
$$ T(u)=\R_{01}(u)\R_{02}(u)....\R_{0L}(u)$$
where we have
\begin{eqnarray}
\R_{0i}(u)&=&R_{0i}(u)~~~~~~~~~~~~~~~~~~~~  \rm{for}~~ i\notin I,   \nonumber \\
\R_{0i}(u)&=&R_{0i'}(u)\oR_{0i''}(u-\lambda)~~~~~~~~  \rm{for}~~ i\in I
\nonumber\end{eqnarray}
and we set $\lambda= \alpha +2$. 
Above, the indices $i'$ and $i''$ refer to the two spaces in $X_i$ (cf.
\reff{w}).
A consequence of eqs. (\ref{yb}, \ref{yb1}) is that the monodromy
matrix satisfies the intertwining relation
\be R_{12}(u-v)T_{13}(u)T_{23}(v)=T_{23}(v)T_{13}(u)R_{12}(u-v) \label{int}
\ee
acting on the space $V\otimes V\otimes X$. The transfer matrix is defined by
\be
\tau(u)=\rm{tr}_0 \sigma_0T(u) \label{tm}
\ee
where the matrix $\sigma$ has entries
$$\sigma_j^i=(-1)^{[i][j]}\delta_j^i$$
from which the Hamiltonian \cite{bef} is obtained through
\be
H=-2\left.\frac d{du}\ln (v^L\tau(u))\right|_{u=0}. \label{ham}
\ee
From \reff{int} it is concluded by usual arguments that the transfer
matrix provides a set of abelian symmetries for the model and hence the
Hamiltonian is integrable. In the next section we will solve the model
by the algebraic Bethe ansatz approach. We use a different grading (BBF) from
that adopted in \cite{bef} (FBB) in order to obtain the Bethe ansatz equations
in a form which is more convenient to prove completeness of the model.
We remark that for the case of one impurity, the equivalence of
these two forms of Bethe ansatz equations has already been discussed in
\cite{bef}. We also present the third form (BFB) of
the Bethe ansatz equations.

\section*{\bf 3. Algebraic Bethe ansatz solution}

By a suitable redefinition of the matrix elements, the
operators $R$ and $R^*$
may be written in terms of matrices which satisfy the Yang-Baxter
equations (\ref{yb}, \ref{yb1})
without ${ \bf Z}_2$-grading (see e.g. \cite{del}). These matrices read
\begin{equation}
\label{r}
{\footnotesize
R^{\gamma \delta}_{\alpha \beta}(v)=
\ba{c}
\unitlength=0.50mm
\begin{picture}(20.,25.)
\put(11.,-4.){\makebox(0.,0.){$\s \beta $}}
\put(24.,11.){\makebox(0.,0.){$\s \alpha $}}
\put(-2.,11.){\makebox(0.,0.){$\s \gamma $}}
\put(11.,24.){\makebox(0.,0.){$\s \delta $}}
\put(0.,11.){\line(1,0){22.}}
\put(11.,21.){\line(0,-1){22.}}
\end{picture}
\ea =
\pmatrix{a&0&0&|&  0&0&0&|&  0&0&0\cr
               0&b&0&|&  c&0&0&|&  0&0&0\cr
               0&0&b&|&  0&0&0&|&  c&0&0\cr
               -&-&-& &  -&-&-& &  -&-&-\cr
               0&c&0&|&  b&0&0&|&  0&0&0\cr
               0&0&0&|&  0&a&0&|&  0&0&0\cr
               0&0&0&|&  0&0&b&|&  0&c&0\cr
               -&-&-& &  -&-&-& &  -&-&-\cr
               0&0&c&|&  0&0&0&|&  b&0&0\cr
               0&0&0&|&  0&0&c&|&  0&b&0\cr
               0&0&0&|&  0&0&0&|&  0&0&w\cr} } \, \, \, ,
\end{equation}
where $\alpha$ , $\beta$ ( $\gamma$ and $\delta$ ) are column ( row
) indices running from 1 to 3 and
\begin{equation}
\label{2}
a =  1 - 2/v ,\quad
b =  1 ,\quad
c = -2/v, \quad
w = -1-2/v \, \, \, .
\end{equation}
and
\begin{equation}
\label{rstar}
{\footnotesize
{R^*}^{\gamma \delta}_{\alpha \beta}(v)=
\ba{c}
\unitlength=0.50mm
\begin{picture}(20.,25.)
\put(11.,-4.){\makebox(0.,0.){$\s \beta $}}
\put(24.,11.){\makebox(0.,0.){$\s \alpha $}}
\put(-2.,11.){\makebox(0.,0.){$\s \gamma $}}
\put(11.,24.){\makebox(0.,0.){$\s \delta $}}
\put(0.,11.){\line(1,0){22.}}
\put(11.00,21.00){\circle{2.0}}
\put(11.00,20.00){\circle{2.0}}
\put(11.00,19.00){\circle{2.0}}
\put(11.00,18.00){\circle{2.0}}
\put(11.00,17.00){\circle{2.0}}
\put(11.00,16.00){\circle{2.0}}
\put(11.00,15.00){\circle{2.0}}
\put(11.00,14.00){\circle{2.0}}
\put(11.00,13.00){\circle{2.0}}
\put(11.00,12.00){\circle{2.0}}
\put(11.00,11.00){\circle{2.0}}
\put(11.00,10.00){\circle{2.0}}
\put(11.00,9.00){\circle{2.0}}
\put(11.00,8.00){\circle{2.0}}
\put(11.00,7.00){\circle{2.0}}
\put(11.00,6.00){\circle{2.0}}
\put(11.00,5.00){\circle{2.0}}
\put(11.00,4.00){\circle{2.0}}
\put(11.00,3.00){\circle{2.0}}
\put(11.00,2.00){\circle{2.0}}
\put(11.00,1.00){\circle{2.0}}
\put(11.00,0.00){\circle{2.0}}

\end{picture}
\ea =
\pmatrix{1&0&0&0&|&  0&0&0&0&|&  0&0&0&0\cr
 0&1&0&0&|&0&0&0&0&|&  0&0&0&0\cr
 0&0&b^*&0&|&  0&c^*&0&0&|& d&0&0&0\cr
 0&0&0&b^*&|&  0&0&0&0&|& 0&e^*&0&0\cr
 -&-&-&-& &  -&-&-&-& &  -&-&-&-\cr
 0&0&0&0&|&a&0&0&0&|&  0&0&0&0\cr
 0&0&c^*&0&|&  0&b^*&0&0&|&-d^*&0&0&0\cr
 0&0&0&0&|&  0&0&a&0&|&  0&0&0&0\cr
 0&0&0&0&|&  0&0&0&b^*&|&  0&0&e^*&0\cr
 -&-&-&-& &  -&-&-&-& &  -&-&-&-\cr
 0&0&d^*&0&|&  0&-d^*&0&0&|&w^*&0&0&0\cr
 0&0&0&-e^*&|&  0&0&0&0&|&  0&-f^*&0&0\cr
 0&0&0&0&|&  0&0&0&-e^*&|&  0&0&-f^*&0\cr
 0&0&0&0&|&  0&0&0&0&|&  0&0&0&g^*\cr} } ,
\end{equation}
with
\begin{eqnarray}
\label{4}
b^* &=& 1 + 2/v' ,\quad
c^* = -2/v' ,\quad
d^* = 2 \sqrt{\alpha}/v'  ,\quad
e^* = - 2 \sqrt{\alpha + 1}/v'  ,\quad \nonumber \\
f^* &=& 1 + 2 (\alpha + 1)/v' ,\quad
w^* = 1 + 2 \alpha /v' , \quad
g^* = 1 + 2 ( \alpha + 2 )/v'   
\label{elem}
\end{eqnarray}
and $v'=v-\alpha-2$. 
The $R$-matrix~(\ref{r}) acts in the tensor product of the two
3-dimensional spaces $V \otimes V$ while
$R^*$~(\ref{rstar}) acts in the tensor product of
$V \otimes W$, where $W$ is the 4-dimensional space.
Hereafter we will use these forms.

Next we construct the monodromy matrix
\begin{equation}
Y(u) = \pmatrix{Y^1_1(u) & Y^2_1(u)  & Y^3_1(u)\cr
                 Y^1_2(u) & Y^2_2(u)  & Y^3_2(u)\cr
                 Y^1_3(u) & Y^2_3(u)  & Y^3_3(u)\cr} \, \, \, ,
\label{monodromy}
\end{equation}
whose elements are operators acting in the quantum space $X$.
The action of these elements is given by
\be
\pi(Y^i_j(u))^{\{ \beta \}}_{\{ \beta^\prime \}}=(-1)^{([i]
[{\{ \beta^\prime \}}]+[j][{\{ \beta^\prime \}}]+[i][{\{ \beta \}}])}
T^{i {\{ \beta \}}}_{j {\{ \beta^\prime \}}}(u),
\label{rep}
\ee
such that $Y(u)$
satisfies the Yang-Baxter algebra without $\bf{Z}_2$-grading
\be
R_{12}(u-v)Y_{13}(u)Y_{23}(v)=Y_{23}(v)Y_{13}(u)R_{12}(u-v) \label{y}
\ee

The transfer matrix is expressible in terms of this
representation by
\begin{equation}
\tau(u)=\sum_{i=1}^3(-1)^{[i]+[i][{\{ \beta \}}]}
\pi(Y^i_i(u))^{\{ \beta \}}_{\{ \beta^\prime \}}
\label{tau}
\end{equation}
In the following we will omit
the symbol $\pi$ for ease of notation.

For a given $\{\alpha\}=(\alpha_1,\alpha_2,...,\alpha_l)$,
$\alpha_i=a,b$ we define the first pseudovacuum $v^{\{\alpha\}}\in X$ by
$$v^{\{\alpha\}}=\bigotimes_{i=1}^L w^i  $$
where
\begin{eqnarray}
w^i=&v^1 ~~~~~~~~~~~&\rm{for}~~ i\notin I,   \nonumber \\
w^i=&v^1\otimes v^{\alpha_j} ~~~&\rm{for}~~ i=p_j\in I. \nonumber
\end{eqnarray}
An example for the case $L=10$ and $l=2$ is depicted below

\[
\unitlength=0.50mm
\begin{picture}(20.,25.)
\put(-70.,5.){\makebox(0.,0.){$v^{\left\{\alpha\right\}}$}}
\put(-50.,0.){\makebox(0.,0.){$=$}}
\put(-40.,-12.){\line(0,1){25.}}
\put(-40.,-20.){\makebox(0.,0.){$1$}}
\put(-30.,-12.){\line(0,1){25.}}
\put(-30.,-20.){\makebox(0.,0.){$1$}}
\put(-20.00,-12.00){\circle{2.0}}
\put(-20.00,-11.00){\circle{2.0}}
\put(-20.00,-10.00){\circle{2.0}}
\put(-20.00,-9.00){\circle{2.0}}
\put(-20.00,-8.00){\circle{2.0}}
\put(-20.00,-7.00){\circle{2.0}}
\put(-20.00,-6.00){\circle{2.0}}
\put(-20.00,-5.00){\circle{2.0}}
\put(-20.00,-4.00){\circle{2.0}}
\put(-20.00,-3.00){\circle{2.0}}
\put(-20.00,-2.00){\circle{2.0}}
\put(-20.00,-1.00){\circle{2.0}}
\put(-20.00,0.00){\circle{2.0}}
\put(-20.00,1.00){\circle{2.0}}
\put(-20.00,2.00){\circle{2.0}}
\put(-20.00,3.00){\circle{2.0}}
\put(-20.00,4.00){\circle{2.0}}
\put(-20.00,5.00){\circle{2.0}}
\put(-20.00,6.00){\circle{2.0}}
\put(-20.00,7.00){\circle{2.0}}
\put(-20.00,8.00){\circle{2.0}}
\put(-20.00,9.00){\circle{2.0}}
\put(-20.00,10.00){\circle{2.0}}
\put(-20.00,11.00){\circle{2.0}}
\put(-20.00,12.00){\circle{2.0}}
\put(-20.00,13.00){\circle{2.0}}
\put(-20.,-20.){\makebox(0.,0.){$\alpha_{1}$}}
\put(-10.,-12.){\line(0,1){25.}}
\put(-10.,-20.){\makebox(0.,0.){$1$}}
\put(0.,-12.){\line(0,1){25.}}
\put(0.,-20.){\makebox(0.,0.){$1$}}
\put(10.,-12.){\line(0,1){25.}}
\put(10.,-20.){\makebox(0.,0.){$1$}}
\put(20.,-12.){\line(0,1){25.}}
\put(20.,-20.){\makebox(0.,0.){$1$}}
\put(30.00,-12.00){\circle{2.0}}
\put(30.00,-11.00){\circle{2.0}}
\put(30.00,-10.00){\circle{2.0}}
\put(30.00,-9.00){\circle{2.0}}
\put(30.00,-8.00){\circle{2.0}}
\put(30.00,-7.00){\circle{2.0}}
\put(30.00,-6.00){\circle{2.0}}
\put(30.00,-5.00){\circle{2.0}}
\put(30.00,-4.00){\circle{2.0}}
\put(30.00,-3.00){\circle{2.0}}
\put(30.00,-2.00){\circle{2.0}}
\put(30.00,-1.00){\circle{2.0}}
\put(30.00,0.00){\circle{2.0}}
\put(30.00,1.00){\circle{2.0}}
\put(30.00,2.00){\circle{2.0}}
\put(30.00,3.00){\circle{2.0}}
\put(30.00,4.00){\circle{2.0}}
\put(30.00,5.00){\circle{2.0}}
\put(30.00,6.00){\circle{2.0}}
\put(30.00,7.00){\circle{2.0}}
\put(30.00,8.00){\circle{2.0}}
\put(30.00,9.00){\circle{2.0}}
\put(30.00,10.00){\circle{2.0}}
\put(30.00,11.00){\circle{2.0}}
\put(30.00,12.00){\circle{2.0}}
\put(30.00,13.00){\circle{2.0}}
\put(30.,-20.){\makebox(0.,0.){$\alpha_{2}$}}
\put(40.,-12.){\line(0,1){25.}}
\put(40.,-20.){\makebox(0.,0.){$1$}}
\put(50.,-12.){\line(0,1){25.}}
\put(50.,-20.){\makebox(0.,0.){$1$}}
\put(60.,-12.){\line(0,1){25.}}
\put(60.,-20.){\makebox(0.,0.){$1$}}
\put(70.,-12.){\line(0,1){25.}}
\put(70.,-20.){\makebox(0.,0.){$1$}}
\put(120.,0.){\makebox(0.,0.){$\alpha_i=a,b$}}
\end{picture}
\]
\vspace{0.50cm}

Setting
$$S^{\{\beta\}}(\{u\})=Y^{\beta_1}_1(u_1)Y^{\beta_2}_1(u_2)....
Y^{\beta_N}_1(u_N),~~~~\beta_i=2,3 $$
we look for a set of eigenstates of the transfer matrix of the form
\be \Phi^j=\sum_{\{\beta,\alpha\}}S^{\{\beta\}}(\{u\})v^{\{\alpha\}}
F^j_{\{\beta,\alpha\}} \label{eig} \ee
where the $F^j_{\{\beta,\alpha\}}$ are undetermined
co-efficients. We appeal to the algebraic equations given by \reff{y} to
determine the constraints on the variables $u_i$
needed to force \reff{eig} to be an
eigenstate.  Although many relations occur as a result of \reff{y} only
the following are required:
\begin{eqnarray}
Y^1_1(v)Y^{\beta}_1(u)&=&a(u-v)Y^{\beta}_1(u)Y^1_1(v)-b(u-v)T_1^{\beta}(v)Y^1_1
(u)  \label{yba1} \\
Y^{\gamma'}_{\gamma}(v)Y^{\alpha}_1(u)&=&Y^{\alpha'}_1(u)Y^{\gamma''}_{
\gamma}(v)r^{\gamma'\alpha}_{\gamma''\alpha'}(v-u)
-b(v-u)Y^{\gamma'}_1(v)Y^{\alpha}_{\gamma}(u)  \label{yba2}
 \\  a(v-u)Y_1^{\alpha}(v)Y_1^{\beta}(u)&=&
Y_1^{\beta'}(u)Y_1^{\alpha'}(v) r^{\beta\alpha}_{\beta'\alpha'}(v-u)
\label{yba3}
\end{eqnarray}
All of the indices in
eqs. (\ref{yba1}, \ref{yba2}, \ref{yba3}) assume only the
values 2 and 3.
Above, the matrix $r(u)$ arises as a submatrix of $R(u)$ and
satisfies the Yang-Baxter equation for a $gl(1|1)$ invariant model.

Using \reff{yba1} two types of terms arise when $Y^1_1$
is commuted through $Y^{\alpha}_1$. In the first type $Y^1_1$ and
$Y_1^{\alpha}$ preserve their arguments and in the second type their
arguments are exchanged. The first type of terms are called {\it wanted
terms} because they will give a vector proportional to $\Phi^j$, and the
second type are {\it unwanted terms} (u.t.).
We find that
\be
Y^1_1(v)\Phi^j=a(v)^L\prod_{i=1}^N a(u_i-v)\Phi^j + \rm{u.t.},
\label{y11} \label{can1}
\ee
where we have used the fact that
\be
Y^1_1(v) v^{\{ \alpha \}}=a(v)^L v^{\{ \alpha \}}
\ee
For the case of the previous example we have 
\[
\unitlength=0.50mm
\begin{picture}(20.,25.)
\put(-45.,0.){\line(1,0){120.}}
\put(-80.,5.){\makebox(0.,0.){$Y_{1}^{1}v^{\left\{\alpha\right\}}$}}
\put(-60.,0.){\makebox(0.,0.){$=$}}
\put(-50.,0.){\makebox(0.,0.){$1$}}
\put(-40.,-12.){\line(0,1){25.}}
\put(-40.,-20.){\makebox(0.,0.){$1$}}
\put(-40.,20.){\makebox(0.,0.){$1$}}
\put(-30.,-12.){\line(0,1){25.}}
\put(-30.,-20.){\makebox(0.,0.){$1$}}
\put(-30.,20.){\makebox(0.,0.){$1$}}
\put(-20.00,-12.00){\circle{2.0}}
\put(-20.00,-11.00){\circle{2.0}}
\put(-20.00,-10.00){\circle{2.0}}
\put(-20.00,-9.00){\circle{2.0}}
\put(-20.00,-8.00){\circle{2.0}}
\put(-20.00,-7.00){\circle{2.0}}
\put(-20.00,-6.00){\circle{2.0}}
\put(-20.00,-5.00){\circle{2.0}}
\put(-20.00,-4.00){\circle{2.0}}
\put(-20.00,-3.00){\circle{2.0}}
\put(-20.00,-2.00){\circle{2.0}}
\put(-20.00,-1.00){\circle{2.0}}
\put(-20.00,0.00){\circle{2.0}}
\put(-20.00,1.00){\circle{2.0}}
\put(-20.00,2.00){\circle{2.0}}
\put(-20.00,3.00){\circle{2.0}}
\put(-20.00,4.00){\circle{2.0}}
\put(-20.00,5.00){\circle{2.0}}
\put(-20.00,6.00){\circle{2.0}}
\put(-20.00,7.00){\circle{2.0}}
\put(-20.00,8.00){\circle{2.0}}
\put(-20.00,9.00){\circle{2.0}}
\put(-20.00,10.00){\circle{2.0}}
\put(-20.00,11.00){\circle{2.0}}
\put(-20.00,12.00){\circle{2.0}}
\put(-20.00,13.00){\circle{2.0}}
\put(-20.,-20.){\makebox(0.,0.){$\alpha_{1}$}}
\put(-20.,20.){\makebox(0.,0.){$\alpha_{1}$}}
\put(-10.,-12.){\line(0,1){25.}}
\put(-10.,-20.){\makebox(0.,0.){$1$}}
\put(-10.,20.){\makebox(0.,0.){$1$}}
\put(0.,-12.){\line(0,1){25.}}
\put(0.,-20.){\makebox(0.,0.){$1$}}
\put(0.,20.){\makebox(0.,0.){$1$}}
\put(10.,-12.){\line(0,1){25.}}
\put(10.,-20.){\makebox(0.,0.){$1$}}
\put(10.,20.){\makebox(0.,0.){$1$}}
\put(20.,-12.){\line(0,1){25.}}
\put(20.,-20.){\makebox(0.,0.){$1$}}
\put(20.,20.){\makebox(0.,0.){$1$}}
\put(30.00,-12.00){\circle{2.0}}
\put(30.00,-11.00){\circle{2.0}}
\put(30.00,-10.00){\circle{2.0}}
\put(30.00,-9.00){\circle{2.0}}
\put(30.00,-8.00){\circle{2.0}}
\put(30.00,-7.00){\circle{2.0}}
\put(30.00,-6.00){\circle{2.0}}
\put(30.00,-5.00){\circle{2.0}}
\put(30.00,-4.00){\circle{2.0}}
\put(30.00,-3.00){\circle{2.0}}
\put(30.00,-2.00){\circle{2.0}}
\put(30.00,-1.00){\circle{2.0}}
\put(30.00,0.00){\circle{2.0}}
\put(30.00,1.00){\circle{2.0}}
\put(30.00,2.00){\circle{2.0}}
\put(30.00,3.00){\circle{2.0}}
\put(30.00,4.00){\circle{2.0}}
\put(30.00,5.00){\circle{2.0}}
\put(30.00,6.00){\circle{2.0}}
\put(30.00,7.00){\circle{2.0}}
\put(30.00,8.00){\circle{2.0}}
\put(30.00,9.00){\circle{2.0}}
\put(30.00,10.00){\circle{2.0}}
\put(30.00,11.00){\circle{2.0}}
\put(30.00,12.00){\circle{2.0}}
\put(30.00,13.00){\circle{2.0}}
\put(30.,-20.){\makebox(0.,0.){$\alpha_{2}$}}
\put(30.,20.){\makebox(0.,0.){$\alpha_{2}$}}
\put(40.,-12.){\line(0,1){25.}}
\put(40.,-20.){\makebox(0.,0.){$1$}}
\put(40.,20.){\makebox(0.,0.){$1$}}
\put(50.,-12.){\line(0,1){25.}}
\put(50.,-20.){\makebox(0.,0.){$1$}}
\put(50.,20.){\makebox(0.,0.){$1$}}
\put(60.,-12.){\line(0,1){25.}}
\put(60.,-20.){\makebox(0.,0.){$1$}}
\put(60.,20.){\makebox(0.,0.){$1$}}
\put(70.,-12.){\line(0,1){25.}}
\put(70.,-20.){\makebox(0.,0.){$1$}}
\put(70.,20.){\makebox(0.,0.){$1$}}
\put(80.,0.){\makebox(0.,0.){$1$}}
\put(120.,0.){\makebox(0.,0.){$\alpha_i=a,b$}}
\end{picture}
\]
\vspace{0.50cm}

Similarly, for $i=2,3$ we have from \reff{yba2} (no sum on $i$)
\begin{eqnarray}
Y^i_i(v)\Phi^j&=&S^{\{\beta'\}}(\{u\})Y^i_k(v)t^{k\{\beta\}}_{i\{\beta'\}}
(v,\{u\})v^{\{\alpha\}}F^j_{\{\beta,\alpha\}} + \rm{u.t.} \nonumber
\\
&=& S^{\{\beta'\}}(\{u\})
t^{k\{\beta\}}_{i\{\beta'\}}(v,\{u\})t^{*i\{\alpha\}}_{\,k\{\alpha'\}}
(v)v^{\{\alpha'\}}F^j_{\{\beta,\alpha\}}
+\rm{u.t} \nonumber \\
&=&
S^{\{\beta'\}}(\{u\})
\overline{t}^{i\{\beta,\alpha\}}_{i\{\beta',\alpha'\}}(v,\{u\})
v^{\{\alpha'\}}F^j_{\{\beta,\alpha\}}   +\rm{u.t.}
\nonumber\end{eqnarray}
where
$$\overline{t}^{i\{\beta,\alpha\}}_{i\{\beta',\alpha'\}}(v,\{u \})
=t^{k\{\beta\}}_{i\{\beta'\}}(v,\{u\})t^{*i\{\alpha\}}_{\,k\{\alpha'\}}
(v). $$
Above
\begin{eqnarray}
t(v,\{u\})&=&r_{01}(v-u_1)r_{02}(v-u_2)...r_{0N}(v-u_N),    \nonumber \\
t^*(v)&=&r^*_{01}(v)r^*_{02}(v)...r^*_{0l}(v).  \nonumber
\end{eqnarray}
and $r^*(u)$ is the $gl(1|1)$ $R$-matrix defined from $R^*(u)$
by restricting the indices to $\alpha, \gamma = 2, 3$ and
$\beta, \delta = a, b $ in (\ref{rstar})
in analogy to $r(u)$.
It is important to observe
that the space $X$ is closed under the action of the elements
$Y^i_j(u),~i,j=2,3$ which generate a $gl(1|1)$-Yang-Baxter algebra.
In fact
$$Y^i_j(u)v^{\{\alpha\}}=t^{*i\{\alpha\}}_{\,j\{\alpha'\}}(u)
v^{\{\alpha'\}}  $$
which follows from the fact that the $Y^i_j(u)~i,j=2,3$ act trivially on
the vector $v^1$ in the sense
\begin{eqnarray}
&&Y^2_2(u)v^1=Y^3_3(u)v^1=v^1 \nonumber \\
&&Y^2_3(u)v^1=Y^3_2(u)v^1=0\nonumber  \end{eqnarray}
An example of the action of $Y^2_2(u)$ and  $Y^3_3(u)$
on the first pseudovacuum
is illustrated below
\vspace{0.50cm}
\[
\unitlength=0.50mm
\begin{picture}(20.,25.)
\put(-45.,0.){\line(1,0){120.}}
\put(-80.,5.){\makebox(0.,0.){$Y_{2}^{2}v^{\left\{\alpha\right\}}$}}
\put(-60.,0.){\makebox(0.,0.){$=$}}
\put(-50.,0.){\makebox(0.,0.){$2$}}
\put(-40.,-12.){\line(0,1){25.}}
\put(-40.,-20.){\makebox(0.,0.){$1$}}
\put(-40.,20.){\makebox(0.,0.){$1$}}
\put(-30.,-12.){\line(0,1){25.}}
\put(-30.,-20.){\makebox(0.,0.){$1$}}
\put(-30.,20.){\makebox(0.,0.){$1$}}
\put(-20.00,-12.00){\circle{2.0}}
\put(-20.00,-11.00){\circle{2.0}}
\put(-20.00,-10.00){\circle{2.0}}
\put(-20.00,-9.00){\circle{2.0}}
\put(-20.00,-8.00){\circle{2.0}}
\put(-20.00,-7.00){\circle{2.0}}
\put(-20.00,-6.00){\circle{2.0}}
\put(-20.00,-5.00){\circle{2.0}}
\put(-20.00,-4.00){\circle{2.0}}
\put(-20.00,-3.00){\circle{2.0}}
\put(-20.00,-2.00){\circle{2.0}}
\put(-20.00,-1.00){\circle{2.0}}
\put(-20.00,0.00){\circle{2.0}}
\put(-20.00,1.00){\circle{2.0}}
\put(-20.00,2.00){\circle{2.0}}
\put(-20.00,3.00){\circle{2.0}}
\put(-20.00,4.00){\circle{2.0}}
\put(-20.00,5.00){\circle{2.0}}
\put(-20.00,6.00){\circle{2.0}}
\put(-20.00,7.00){\circle{2.0}}
\put(-20.00,8.00){\circle{2.0}}
\put(-20.00,9.00){\circle{2.0}}
\put(-20.00,10.00){\circle{2.0}}
\put(-20.00,11.00){\circle{2.0}}
\put(-20.00,12.00){\circle{2.0}}
\put(-20.00,13.00){\circle{2.0}}
\put(-20.,-20.){\makebox(0.,0.){$\alpha_{1}$}}
\put(-20.,20.){\makebox(0.,0.){$\alpha_{1}$}}
\put(-10.,-12.){\line(0,1){25.}}
\put(-10.,-20.){\makebox(0.,0.){$1$}}
\put(-10.,20.){\makebox(0.,0.){$1$}}
\put(0.,-12.){\line(0,1){25.}}
\put(0.,-20.){\makebox(0.,0.){$1$}}
\put(0.,20.){\makebox(0.,0.){$1$}}
\put(10.,-12.){\line(0,1){25.}}
\put(10.,-20.){\makebox(0.,0.){$1$}}
\put(10.,20.){\makebox(0.,0.){$1$}}
\put(20.,-12.){\line(0,1){25.}}
\put(20.,-20.){\makebox(0.,0.){$1$}}
\put(20.,20.){\makebox(0.,0.){$1$}}
\put(30.00,-12.00){\circle{2.0}}
\put(30.00,-11.00){\circle{2.0}}
\put(30.00,-10.00){\circle{2.0}}
\put(30.00,-9.00){\circle{2.0}}
\put(30.00,-8.00){\circle{2.0}}
\put(30.00,-7.00){\circle{2.0}}
\put(30.00,-6.00){\circle{2.0}}
\put(30.00,-5.00){\circle{2.0}}
\put(30.00,-4.00){\circle{2.0}}
\put(30.00,-3.00){\circle{2.0}}
\put(30.00,-2.00){\circle{2.0}}
\put(30.00,-1.00){\circle{2.0}}
\put(30.00,0.00){\circle{2.0}}
\put(30.00,1.00){\circle{2.0}}
\put(30.00,2.00){\circle{2.0}}
\put(30.00,3.00){\circle{2.0}}
\put(30.00,4.00){\circle{2.0}}
\put(30.00,5.00){\circle{2.0}}
\put(30.00,6.00){\circle{2.0}}
\put(30.00,7.00){\circle{2.0}}
\put(30.00,8.00){\circle{2.0}}
\put(30.00,9.00){\circle{2.0}}
\put(30.00,10.00){\circle{2.0}}
\put(30.00,11.00){\circle{2.0}}
\put(30.00,12.00){\circle{2.0}}
\put(30.00,13.00){\circle{2.0}}
\put(30.,-20.){\makebox(0.,0.){$\alpha_{2}$}}
\put(30.,20.){\makebox(0.,0.){$\alpha_{2}$}}
\put(40.,-12.){\line(0,1){25.}}
\put(40.,-20.){\makebox(0.,0.){$1$}}
\put(40.,20.){\makebox(0.,0.){$1$}}
\put(50.,-12.){\line(0,1){25.}}
\put(50.,-20.){\makebox(0.,0.){$1$}}
\put(50.,20.){\makebox(0.,0.){$1$}}
\put(60.,-12.){\line(0,1){25.}}
\put(60.,-20.){\makebox(0.,0.){$1$}}
\put(60.,20.){\makebox(0.,0.){$1$}}
\put(70.,-12.){\line(0,1){25.}}
\put(70.,-20.){\makebox(0.,0.){$1$}}
\put(70.,20.){\makebox(0.,0.){$1$}}
\put(80.,0.){\makebox(0.,0.){$2$}}
\put(120.,0.){\makebox(0.,0.){$\alpha_i=a,b$}}
\end{picture}
\]
\vspace{1.00cm}
\[
\unitlength=0.50mm
\begin{picture}(20.,25.)
\put(-45.,0.){\line(1,0){120.}}
\put(-80.,5.){\makebox(0.,0.){$Y_{3}^{3}v^{\left\{\alpha\right\}}$}}
\put(-60.,0.){\makebox(0.,0.){$=$}}
\put(-50.,0.){\makebox(0.,0.){$3$}}
\put(-40.,-12.){\line(0,1){25.}}
\put(-40.,-20.){\makebox(0.,0.){$1$}}
\put(-40.,20.){\makebox(0.,0.){$1$}}
\put(-30.,-12.){\line(0,1){25.}}
\put(-30.,-20.){\makebox(0.,0.){$1$}}
\put(-30.,20.){\makebox(0.,0.){$1$}}
\put(-20.00,-12.00){\circle{2.0}}
\put(-20.00,-11.00){\circle{2.0}}
\put(-20.00,-10.00){\circle{2.0}}
\put(-20.00,-9.00){\circle{2.0}}
\put(-20.00,-8.00){\circle{2.0}}
\put(-20.00,-7.00){\circle{2.0}}
\put(-20.00,-6.00){\circle{2.0}}
\put(-20.00,-5.00){\circle{2.0}}
\put(-20.00,-4.00){\circle{2.0}}
\put(-20.00,-3.00){\circle{2.0}}
\put(-20.00,-2.00){\circle{2.0}}
\put(-20.00,-1.00){\circle{2.0}}
\put(-20.00,0.00){\circle{2.0}}
\put(-20.00,1.00){\circle{2.0}}
\put(-20.00,2.00){\circle{2.0}}
\put(-20.00,3.00){\circle{2.0}}
\put(-20.00,4.00){\circle{2.0}}
\put(-20.00,5.00){\circle{2.0}}
\put(-20.00,6.00){\circle{2.0}}
\put(-20.00,7.00){\circle{2.0}}
\put(-20.00,8.00){\circle{2.0}}
\put(-20.00,9.00){\circle{2.0}}
\put(-20.00,10.00){\circle{2.0}}
\put(-20.00,11.00){\circle{2.0}}
\put(-20.00,12.00){\circle{2.0}}
\put(-20.00,13.00){\circle{2.0}}
\put(-20.,-20.){\makebox(0.,0.){$\alpha_{1}$}}
\put(-20.,20.){\makebox(0.,0.){$\alpha_{1}$}}
\put(-10.,-12.){\line(0,1){25.}}
\put(-10.,-20.){\makebox(0.,0.){$1$}}
\put(-10.,20.){\makebox(0.,0.){$1$}}
\put(0.,-12.){\line(0,1){25.}}
\put(0.,-20.){\makebox(0.,0.){$1$}}
\put(0.,20.){\makebox(0.,0.){$1$}}
\put(10.,-12.){\line(0,1){25.}}
\put(10.,-20.){\makebox(0.,0.){$1$}}
\put(10.,20.){\makebox(0.,0.){$1$}}
\put(20.,-12.){\line(0,1){25.}}
\put(20.,-20.){\makebox(0.,0.){$1$}}
\put(20.,20.){\makebox(0.,0.){$1$}}
\put(30.00,-12.00){\circle{2.0}}
\put(30.00,-11.00){\circle{2.0}}
\put(30.00,-10.00){\circle{2.0}}
\put(30.00,-9.00){\circle{2.0}}
\put(30.00,-8.00){\circle{2.0}}
\put(30.00,-7.00){\circle{2.0}}
\put(30.00,-6.00){\circle{2.0}}
\put(30.00,-5.00){\circle{2.0}}
\put(30.00,-4.00){\circle{2.0}}
\put(30.00,-3.00){\circle{2.0}}
\put(30.00,-2.00){\circle{2.0}}
\put(30.00,-1.00){\circle{2.0}}
\put(30.00,0.00){\circle{2.0}}
\put(30.00,1.00){\circle{2.0}}
\put(30.00,2.00){\circle{2.0}}
\put(30.00,3.00){\circle{2.0}}
\put(30.00,4.00){\circle{2.0}}
\put(30.00,5.00){\circle{2.0}}
\put(30.00,6.00){\circle{2.0}}
\put(30.00,7.00){\circle{2.0}}
\put(30.00,8.00){\circle{2.0}}
\put(30.00,9.00){\circle{2.0}}
\put(30.00,10.00){\circle{2.0}}
\put(30.00,11.00){\circle{2.0}}
\put(30.00,12.00){\circle{2.0}}
\put(30.00,13.00){\circle{2.0}}
\put(30.,-20.){\makebox(0.,0.){$\alpha_{2}$}}
\put(30.,20.){\makebox(0.,0.){$\alpha_{2}$}}
\put(40.,-12.){\line(0,1){25.}}
\put(40.,-20.){\makebox(0.,0.){$1$}}
\put(40.,20.){\makebox(0.,0.){$1$}}
\put(50.,-12.){\line(0,1){25.}}
\put(50.,-20.){\makebox(0.,0.){$1$}}
\put(50.,20.){\makebox(0.,0.){$1$}}
\put(60.,-12.){\line(0,1){25.}}
\put(60.,-20.){\makebox(0.,0.){$1$}}
\put(60.,20.){\makebox(0.,0.){$1$}}
\put(70.,-12.){\line(0,1){25.}}
\put(70.,-20.){\makebox(0.,0.){$1$}}
\put(70.,20.){\makebox(0.,0.){$1$}}
\put(80.,0.){\makebox(0.,0.){$3$}}
\put(120.,0.){\makebox(0.,0.){$\alpha_i=a,b$}}.   
\end{picture}
\]
\vspace{0.50cm}

The contribution to the eigenvalues of the transfer matrix is
\be Y^2_2(v)\Phi^j+(-1)^{1+[j]}Y^3_3(v)\Phi^j
=\sum_{i=2}^3(-1)^{[i]+[i][j]}\overline{t}^{i\{\beta,\alpha\}}_{i\{\beta',\alpha'\}}(v,\{u \})
S^{\{\beta'\}}(\{u\})v^{\{\alpha'\}}F^j_{\{\beta,\alpha\}}+\rm{u.t.}
\label{can2} \ee
At this point we need to perform a second-level, or {\it nested} Bethe ansatz
procedure to diagonalize the matrix
$$
\tau_1(v)^{\{\beta,\alpha\}}_{\{\beta',\alpha'\}}
=\sum_{i=2}^3(-1)^{[i]+[i][\{\beta,\alpha\}]}\overline{t}^{i\{\beta,\alpha\}}_{i
\{\beta',\alpha'\}}(v,\{u \}) $$
where we have used the fact that $F^j_{\{\beta,\alpha\}}=0$ unless
$[j]=[\{\beta,\alpha\}]$.
The above matrix is simply the transfer matrix for a $gl(1|1)$ invariant
system acting in the tensor product representation of $N$ copies of the vector
representation with inhomogeneities $\{u\}$ and $l$ copies of the impurity
sub-representation.

We diagonalize $\tau_1(v)$ in analogy to
$ \tau(v)$ (\ref{tau}) by constructing the nested monodromy matrix
\begin{equation}
y(u) = \pmatrix{ y^2_2(u)  & y^3_2(u)\cr
                 y^2_3(u)  & y^3_3(u)\cr} \, \, \, ,
\label{nested}
\end{equation}
satisfying
\be r_{12}(u-v)y_{13}(u)y_{23}(v)=y_{23}(v)y_{13}(u)r_{12}(u-v).
\label{yy} \ee
From the above set of relations we will need the following
\begin{eqnarray}
y^2_2(v)y^3_2(u)&=&a(u-v)y^3_2(u)y^2_2(v)-b(u-v)y^3_2(v)y^2_2(u), \label{y1}\\
y^3_3(v)y^3_2(u)&=&-a(u-v)y^3_2(u)y^3_3(v)-b(v-u)y^3_2(v)y^3_3(u),
\label{y2}\\
y^3_2(v)y^3_2(u)&=&\frac{-a(u-v)}{a(v-u)}y^3_2(u)y^3_2(v). \label{y3} \end{eqnarray}
Proceeding similarly as before, we look for eigenstates of the form
$$\phi=y^3_2(\gamma_1)y^3_2(\gamma_2)...y^3_2(\gamma_M) \theta $$
with the second-level pseudovacuum $\theta$ given by
$$\theta=S^{\{2\}}(\{u\})v^{\{a\}}.$$
It is an eigenstate of $y^2_2$ and $y^3_3$
\begin{equation}
y^2_2(v) \theta = \prod_{i=1}^N a(v-u_i) \theta \,  , \quad \quad
y^3_3(v) \theta = {[w^*(v)]}^l \theta \, \, .
\label{nest2}
\end{equation}
Using (\ref{y1},\ref{y2}) it follows that
$$\tau_1(v)\phi=\Lambda_1(v)\phi +\rm{u.t.} $$
with
$$ \Lambda_1(v)=\prod_{i=1}^N a(v-u_i)\prod_{k=1}^M
a(\gamma_k-v) -  {[w^*(v)]}^l \prod_{k=1}^M a(\gamma_k-v)$$
The unwanted terms cancel provided the parameters $\gamma_k$  satisfy
the Bethe ansatz equations (BAE)
\be
\prod_{i=1}^N a(\gamma_k-u_i)= {[w^*(\gamma_k)]}^l,
~~~~k=1,2,...,M.
\label{bae1}
\ee
Combining this result with \reff{y11} we obtain for the eigenvalues of
the transfer matrix  \reff{tm}
\be \Lambda(v)=a(v)^L\prod_{i=1}^Na(u_i-v)+\Lambda_1(v). \label{eigv}\ee
Cancellation of the unwanted terms in (\ref{can1},\ref{can2}) leads to a second set of BAE
which are
\be
a(u_h)^L\prod_{i=1}^N\frac{a(u_i-u_h)}{a(u_h-u_i)}=-
\prod_{k=1}^Ma(\gamma_k-u_h) ~~~~h=1,2,....,N.  \label{bae2}
\ee
Using (\ref{2},\ref{4}) and making the change of variables 
 $$u_{h}\rightarrow  iu_{h}+1$$
 $$\gamma_{k} \rightarrow  i\gamma_{k}+2 $$
we obtain
\begin{equation}
\label{1bae}
\left(\frac{u_{h}+i}{u_{h}-i}\right)^{L}=
\prod_{k\not=h}^{N}\frac{u_{h}-u_{k}+2i}{u_{h}-u_{k}-2i}
\prod_{\delta=1}^{M}\frac{u_{h}-\gamma_{\delta}-i}
{u_{h}-\gamma_{\delta}+i}
\;\;\;\;h=1,\ldots N=L+1-n_{\uparrow}   
\end{equation}

\begin{equation}
\label{2bae}
\prod_{h=1}^{N}\frac{\gamma_{\delta} - u_{h}+
i}{\gamma_{\delta}- u_{h} - i}=
\left(\frac{\gamma_{\delta}-i\alpha}{\gamma_{\delta}+
i \alpha }\right)^{l}
\;\;\;\;\delta=1,\ldots M=L+2-n_{\uparrow} -n_{\downarrow}
\end{equation}
In the absence of impurities (limit $l \longrightarrow$ 0) we recover the form of the BAE first
derived by Sutherland \cite{su} and later by Sarkar \cite{sar} for the usual
$t-J$ model.
The basic procedure to solve (\ref{1bae},\ref{2bae}) is to adopt the string-conjecture,
which means that the $u_h$ appear as strings and all roots
$\gamma_{\delta}$ are real
\begin{eqnarray}
\label{conjec}
u_{\alpha j}^{n}&=&u_{\alpha}^{n}+i\left(n+1-2j\right)\;\;\;;j=1,\ldots,n\;\;\;\;\alpha=1,\ldots ,N_{n}
\;\;\; n=1,2,3,\ldots \nonumber \\
\gamma_{\delta}&=&\rm{real}\;\;\; \delta=1,2,3,\ldots M
\end{eqnarray}
where $u_{\alpha}^{n}$ is the position of the center of the string on the real
$u$-axis. The number of $n$-strings $N_{n}$ satisfy the relation
\[
N=\sum_{n}nN_{n}
\]
Applying the string conjecture (\ref{conjec}) in (\ref{1bae}) and (\ref{2bae}) and
taking the product over $j$ in (\ref{1bae}) and (\ref{2bae}) we obtain
\begin{equation}
\label{t1bae}
e^{L}\left(\frac{ u_{\alpha}^{n}}{n}\right)=
\prod_{m}\prod_{\beta}E_{nm}(u_{\alpha}^{n}-
u_{\beta}^{m})\prod_{\delta}^{M}e^{-1}
\left(\frac{u_{\alpha}^{n}-\gamma_{\delta}}{n}\right)
\end{equation}

\begin{equation}
\label{t2bae}
\prod_{n}\prod_{\alpha} e\left(\frac{ \gamma_{\delta} -
u_{\alpha}^{n}}{n}\right)=
e^{-l}\left(\frac{\gamma_{\delta}}{\alpha}\right)
\end{equation}
where $e(x)=\frac{x+i}{x-i}$ and

\begin{eqnarray}
\label{Enm}
E_{nm}(x)&=& e\left(\frac{x}{|n-m|}\right) e^{2}\left(\frac{x}{|n-m|+2}\right)\ldots
e^{2}\left(\frac{x}{n+m-2}\right) e\left(\frac{x}{n+m}\right),\,\,\, n\not=m \nonumber \\& &
e^{2}\left(\frac{x}{2}\right) e^{2}\left(\frac{x}{4}\right)\ldots e^{2}
\left(\frac{x}{2n-2}\right) e^{2}\left(\frac{x}{2n}\right),\,\,\,\  n=m
\end{eqnarray}
Taking the logarithm of these equation and
using $\ln(e\left(x\right))=\frac{2}{i}\arctan\left(x\right)-
\ln\left(-1\right)$
we get
\begin{equation}
\label{L1bae}
L\theta\left(\frac{u_{\alpha}^{n}}{n}\right)
-\sum_{m}\sum_{\beta=1}^{N_{m}}\Theta_{nm}
\left(u_{\alpha}^{n}-
u_{\beta}^{m}\right)+
\sum_{\delta=1}^{M}\theta\left(\frac{u_{\alpha}^{n}-
\gamma_{\delta}}{n}\right)=
2\pi I_{\alpha}^{n}
\end{equation}

\begin{equation}
\label{L2bae}
\sum_{n} \sum_{\alpha=1}^{N_{n}} \theta\left(
\frac{\gamma_\delta - u_{\alpha}^{n} }{n}
\right)+l\theta(\frac{\gamma_{\delta}}{\alpha})=
2 \pi J_{\delta}
\end{equation}
where $\theta(x)=2\arctan(x)$ and

\begin{eqnarray}
\label{teta}
\Theta_{nm}(x)&=&\theta\left(\frac{x}{|n-m|}\right)+2 \theta
\left(\frac{x}{|n-m|+2}\right)+.....+2 \theta\left(\frac{x}{n+m-2}\right)+
\theta\left(\frac{x}{n+m}\right),\,\,\,\,\,\ n\not=m \nonumber \\ & &
2 \theta\left(\frac{x}{2}\right)+2 \theta\left(\frac{x}{4}\right)
+ \dots +2 \theta\left(\frac{x}{2n-2}\right)+
\theta\left(\frac{x}{2n}\right), \,\,\,\,n=m
\end{eqnarray}

Hence the solutions of (\ref{1bae}) and (\ref{2bae}) are parametrized in term of the 
numbers $I_{\alpha}^{n}$ and $J_{\beta}$.
The numbers $I_{\alpha}^{n}$ and $J_{\beta}$ are limited to the intervals
\begin{equation}
\label{Imax}
|I_{\alpha}^{n}|\leq I_{max}^{n}=\frac{1}{2}\left(L+M-\sum_{m}t_{nm}N_{m}-1\right)
\end{equation}

\begin{equation}
\label{Jmax}
|J_{\delta}|\leq J_{max}=\frac{1}{2}\left(\sum_{n}N_{n}+l-2\right)
\end{equation}
where $t_{nm}=2\rm{min}(n,m)-\delta_{n,m}$.

As was shown in the papers \cite{ek,fk} two other forms of the
Bethe ansatz  exist which are obtained by choosing a different
grading for the indices of the $gl(2|1)$ generators.
Recall that the above calculations were performed with the choice
$$[1]=[2]=0,~~~~[3]=1. $$
Choosing
$$[1]=1,~~~~[2]=[3]=0 $$
yields the eigenvalue expression
\begin{eqnarray}
\Lambda(v)&=& - {[-w(v)]}^L {[g^*(v)]}^l \prod_{i=1}^N a(v-u_i)
\nonumber \\
&+& {[b^*(v)]}^l
\biggl( \prod_{i=1}^N a(v-u_i) \prod_{k=1}^M a(\gamma_k-v)
 + \prod_{k=1}^M a(v-\gamma_k) \biggr)
\end{eqnarray}
subject to the BAE
\begin{eqnarray}
{[-w(u_i)]}^L { \biggl(\frac{ g^*(u_i)}{b^*(u_i)} \biggr)}^l &=&
\prod_{k=1}^M a(\gamma_k-u_i),
~~~~i=1,2,...,N  \nonumber \\
\prod_{k=1}^M\frac{a(\gamma_h-\gamma_k)}{a(\gamma_k-\gamma_h)}
&=&-\prod_{i=1}^N a(\gamma_h-u_i)
,~~~~h=1,2,...,M.\nonumber
\end{eqnarray}
In the limit $l \rightarrow 0$ we recover Lai's form of the
BAE \cite{lai} (see also \cite{schlot}).
Alternatively, choosing
$$[1]=[3]=0,~~~~[2]=1 $$
yields the eigenvalue expression
\begin{eqnarray}
\Lambda(v)&=&a(v)^L\prod_{i=1}^N a(u_i-v)+
{[b^*(v)]}^l \prod_{k=1}^M a(v-\gamma_k) \nonumber \\
&-& {[f^*(v)]}^l \prod_{i=1}^N a(u_i-v) \prod_{k=1}^M
a(v-\gamma_k) \nonumber  \end{eqnarray}
with the BAE
\begin{eqnarray}
a(u_i)^L &=& {[f^*(v)]}^l\prod_{k=1}^M a (u_i-\gamma_k)
~~~~i=1,2,...,N, \nonumber \\
1 &=& {\biggl( \frac{f^*(\gamma_k)}{b^*(\gamma_k)}\biggr) }^l
\prod_{i=1}^N a(u_i-\gamma_k)
~~~~ k=1,2,...,M.
\end{eqnarray}

Finally, from the definition of the Hamiltonian \reff{ham} we see that
the energies are given by
$$E=-2\left.\frac{d}{dv} \ln\left(v^L\Lambda(v)\right)\right|_{v=0}. $$
Using the eigenvalue expression \reff{eigv} we obtain
$$E=L-4\sum_{i=1}^N\frac{1}{1+u_i^2}$$
where the $u_i$ are solutions to the equations (\ref{1bae},\ref{2bae}).

\section*{\bf 4. Highest weight property}

Next we wish to show that the eigenstates constructed in the previous
section are in fact highest weight states with respect to the underlying
supersymmetry algebra $gl(2|1)$. The highest weight property of the
Bethe states has been proved for many models, such as the
Heisenberg chain \cite{fad} and its generalized version \cite{kir2},
the Kondo model \cite{kir1}, the usual $t-J$ model \cite{fk},
the Hubbard chain \cite{fab1} and its $gl(2/2)$ extension
\cite{schout}. In all of these examples the Bethe ansatz was performed
using a unique reference state. However it was shown recently that the
highest weight property can also hold for cases where the Bethe ansatz
solution is obtained from a subspace of reference states \cite{angi5}.

Let us begin by considering
$$E^2_3\Phi^j=\sum_{\{\beta,\alpha\}}E^2_3S^{\{\beta\}}(\{u\})v^{\{\alpha
\}}F^j_{\{\beta,\alpha\}}. $$
By means of the nesting procedure we know that the co-efficients $F^j_{
\{\beta,\alpha\}}$ are such that we have the following
identification of states
$$S^{\{\beta\}}(\{u\})v^{\{\alpha\}}F^j_{\{\beta,\alpha\}}
=y^3_2(\gamma_1)y^3_2(\gamma_2)....y^3_2(\gamma_M)w $$
for a suitable solution of the BAE. By comparing eqs.
(\ref{lop},\ref{rep},\ref{yy}) it is possible to determine algebraic
relations between the elements of the Yangian algebra and the
supersymmetry algebra. For our purposes we need the following
\be [E^2_3,\,y^3_2(u)]^{\alpha}_{\beta}=-y^2_2(u)^{\alpha}_{\beta}+
y^3_3(u)^{\alpha}_{\beta}(-1)^{[\alpha]}. \label{tran1} \ee
Noting that $E^2_3w=0$ it is evident that we may write
$$E^2_3y^3_2(\gamma_1)....y^3_2(\gamma_M)w=\sum_{h=1}^Mx_hX_h $$
with
$$X_h=y^3_2(\gamma_1).....y^3_2(\gamma_{h-1})y^3_2(\gamma_{h+1})....y^3_2
(\gamma_M)w $$
and the $x_h$ some yet to be determined co-efficients.
To find $x_h$ we write
$$y^3_2(\gamma_1)....y^3_2(\gamma_M)w=\prod_{j=1}^{h-1}\frac{-a(\gamma_h-
\gamma_j)}{a(\gamma_j-\gamma_h)} y^3_2(\gamma_h)X_h$$
where we have used \reff{y3}. Now by using the relations
(\ref{y1},\ref{y2},\ref{tran1}) and looking only for those terms which
give a vector proportional to $X_h$ we find that
$$x_h=\prod_{j=1}^{h-1}\frac{-a(\gamma_h-\gamma_j)}{a(\gamma_j-\gamma_h)}
\left([w^*(\gamma_h)]^l\prod_{k\neq h}^Ma(\gamma_k-\gamma_h)
-\prod_{i=1}^Na(\gamma_h-u_i)\prod_{k\neq h}^Ma(\gamma_k-\gamma_h) \right) $$
which vanishes because of \reff{bae1}. Thus we see that
$$E^2_3\Phi^j=0.$$

Next we consider the action of $E^1_2$ on $\Phi^j$.
Using eqs. (\ref{lop},\ref{y},\ref{rep}) we find the
commutation relation
\be [E^1_2,\,Y^{\alpha}_1(u)]=\delta^{\alpha}_2Y^1_1(u)-Y^{\alpha}_2(u).
\label{tran2} \ee
As before, since $E^1_2v^{\{\alpha\}}=0$ we can write the general
expression
$$ E^1_2\Phi^j=\sum_{h,\beta} z_{h,\beta}Z_{h,\beta} $$
where
$$Z_{h,\beta}=S^{\{\beta_h^-\}}(\{u_h^-)\})S^{\{\beta_h^+\}}(\{u_h^+)\}
)v^{\{\alpha\}}F^j_{\{\beta,\alpha\}}    $$
and for any vector $\{w\}$ we have
$$\{w_h^-\}=(w_1,w_2,...,w_{h-1}),~~\{w_h^+\}=( w_{h+1},.....,w_N).
$$
To calculate $z_{h,\beta}$ we begin by writing
\begin{eqnarray}
\Phi^j&=&S^{\{\beta_h^-\}}(\{u_h^-\})Y^{\beta_h}_1(u_h)S^{\{\beta_h^+\}}(\{u_h^+\})v^{\{\alpha\}}F^j_{\{\beta,\alpha\}}  \nonumber \\
&=&\prod_{i=1}^{h-1}a(u_i-u_h)^{-1}t^{\beta_h\{\beta_h^-\}}_{
\gamma\{\gamma_h^-\} }(-u_h,\{-u_h^-\})Y^{
\gamma}_1(u_h)S^{\{\gamma_h^-\}}(\{u_h^-\})S^{\{\beta_h^+\}}(\{u_h^+\}
)v^{\{\alpha\}}F^j_{\{\beta,\alpha\}}
\nonumber \end{eqnarray}
where we have used the relation \reff{yba3}.
Now applying \reff{tran2} and using the relations (\ref{yba1},\ref{yba2})
to determine the terms which
give a vector proportional to $Z_{h,\beta}$ we find that
$$z_{h,\beta}=\delta_2^{\beta_h}\left(a(u_h)^L\prod_{i\neq
h}^Na(u_i-u_h) -\prod_{i\neq
h}^Na(u_h-u_i)\prod_{k=1}^Ma(\gamma_k-u_h)\right) $$
which vanishes as a result of \reff{bae2}. We then conclude that
$$E^1_2\Phi^j=0.$$
Finally, using the fact that
$$E^1_3=[E^1_2,\,E^2_3] $$
it is deduced that
$$E^1_3\Phi^j=0$$
which completes the proof that the Bethe states are $gl(2|1)$ highest
weight states. We observe that this property can also be proved
for the other two choices of gradings in a similar way.

By acting with the $gl(2|1)$ lowering operators on the Bethe states
additional states are obtained which are necessarily eigenstates of the
Hamiltonian as a consequence of the $gl(2|1)$ invariance. Before
proceeding to counting the number of eigenstaes, it is important to
assert that each of the Bethe states belongs to an
{\it irreducible} submodule of the full tensor product space.
Recall that unlike the case of Lie algebras, there exist
representations of Lie superalgebras which can be indecomposable but not
irreducible \cite{snr}. However this situation does not occur in the
decomposition of the tensor representation here, which
follows from the fact that both the fundamental representation and the
family of four dimensional representations with $\alpha >0$ belong to
the class of type I unitary (or star) representations defined in \cite{snr} and
classified in \cite{gz}. A general theorem states \cite{snr}
that tensor products
of type I unitary representations are always completely reducible into
representations which are also of type I. Moreover, since the module
which acts for the impurity spaces is typical it follows from \cite{gl}
(see Proposition 2) that all the modules occuring in the decomposition
of the tensor product space are in fact typical.

\section*{\bf 5. Completeness of the Bethe states}
In this section we show  how to construct a complete set of eigenvectors 
for the
$t-J$ hamiltonian for an arbitary chain of length $L$ with $l$-impurities.
This is 
obtained by combining the Bethe ansatz equations with the supersymmetry 
of the model.

From the previous section we know that all collections $\{I^n_\alpha,
J_\delta \}$ where the $I$'s and $J$'s are pairwise
different specify all the Bethe vectors.
The number of admissible values for the $I^n_\alpha$'s and the
$J_\delta $'s (for fixed $\{N_n\}$ and $M$) is $(2 I^n_{\rm max} +1)$ and
$(2 J_{\rm max} + 1)$, respectively. 
 Taking into account that many
different string configurations $N_n$ give the same number of roots $N$
, the number of possible Bethe vectors for fixed $N,M$
is given by
\begin{equation}
Z(N,M) = \sum_{ { \{ N_n\} } } {2J_{\rm max}+1 \choose M } \prod_n
 { 2I^n_{\rm max}+1 \choose N_n} ,
\label{6.1}
\end{equation}
where the sum over $ \{N_n\}$ is constrained to $\sum_n nN_n = N $.
 It is convenient to introduce the quantity $ q =\sum_n N_n $.
 Then we can write this sum as follows
\begin{equation}
Z(N,M) = \sum_{q=0}^N { q+l - 1 \choose M } \sum_{ \{N_n\} }
         \prod_n { L - \sum_m t_{nm} N_m + M \choose N_n},
\label{6.2}
\end{equation}
where the inner sum is constrained to fixed values of $N$ and
$q$.
This expression resembles the one calculated by Bethe in the isotropic
Heisenberg model \cite{hb,fad} and can be simplified to
\begin{equation}
Z(N,M) = \sum_{q=0}^N { L + M - 2N + 1 \over L + M - N + 1 }   { q+l-1
\choose M }   { L + M - N + 1 \choose q }
   { N - 1 \choose q - 1 } .
\label{6.3}
\end{equation}
The total number of Bethe vectors is obtained
by summing $Z(N,M)$ over all $N,M$.
From the fact that the Bethe vectors are highest weight vectors, by
acting with the $gl(2|1)$ lowering operators $E^{i}_{j}\left(i\geq j\right)$
 on the Bethe states we obtain additional states.
Each Bethe state (with fixed $N,M$) is the
highest weight vector in an irreducible multiplet of dimension \cite{snr}
$$d(N,M) =8 ( S_z + 1/2 ) = 4 ( L - 2N + M + 1 )$$
Notice that here, in constrast to the usual $t-J$ model \cite{fk} only
typical $gl(2|1)$ multiplets occur as discussed in the previous section.
With these considerations, the total number of eigenvectors is
$$
  Z = \sum_{M=0}^{L+2l} \sum_{N=M-l}^{ {L + M \over 2} } d(N,M) Z(N,M).$$
Above the ranges of $M$ and $N$ are obtained by demanding that the magnetization
$S_{z}=\frac{1}{2}\left(n_{\uparrow}-n_{\downarrow}\right)=\frac{1}{2}\left(L-2N+M\right)$
and the number of electron $Q=n_{\uparrow}+n_{\downarrow}=L+2l-M$ are restricted
to intervals $0\leq S_{z}\leq \frac{L+l}{2}$ and $0\leq Q \leq L+2l$.
\begin{eqnarray}
\label{c1}
Z&=&\sum_{M=0}^{L+2l}\sum_{N=M-l}^{L+M\over 2}4\left(L-2N+M+1\right) {L+M-2M+1\over
L+M-N+1}\times \nonumber \\& &
\sum_{q=0}^{N} {q+l-1\choose M} {L+M-N+1\choose q}{N-1\choose q-1}
\end{eqnarray}
Using  the identity
$${q+l-1 \choose M}=\sum_{k=0}^{l-1} {l-1 \choose k} {q \choose M-k}\;\;\;\;\;l\geq 1\;\;\;\;,$$
eq.(\ref{c1}) can be written as
\begin{eqnarray}
\label{c2}
Z&=&\sum_{k=0}^{l-1}{l-1 \choose k}\sum_{M=0}^{L+2l}\sum_{N=M-l}^{L+M\over 2}4\left(L-2N+M+1\right) {L+M-2M+1\over
L+M-N+1}\times \nonumber \\ & &\sum_{q=0}^{N} {q\choose M-k} {L+M-N+1\choose q}{N-1\choose q-1} \;\;\;\;.
\end{eqnarray}
Employing some combinatorics, we find
\begin{eqnarray}
\label{c3}
Z&=&\sum_{k=0}^{l-1}{l-1 \choose k}\sum_{M=0}^{L+2l}\sum_{N=M-l}^{L+M\over 2}4\left(L-2N+M+1\right)
\times \nonumber \\ & & {L+M-2N+1\over L+M-N+1} {L+M-N+1\choose M-k}{L+k\choose N-M+k}\;\;\;\;.
\end{eqnarray}
After rearrangement this expression can be rewritten as
\begin{eqnarray}
\label{c4}
Z&=&\sum_{k=0}^{l-1}{l-1 \choose k}\sum_{M=0}^{L+2l}\sum_{N=M-l}^{L+M\over 2}4\left(L-2N+M+1\right)
\times \nonumber \\ & &\left\{{L+M-N+1 \choose M-k} {L+k\choose N-M+k}-{N\choose M-k}{L+k \choose N-1}\right\}\;\;\;\;.
\end{eqnarray}
Making the substitution $N\to x+M-l$ we obtain
\begin{eqnarray}
\label{c5}
Z&=&\sum_{k=0}^{l-1}{l-1 \choose k}\sum_{M=0}^{L+2l}\sum_{x=0}^{{L-M\over 2}+l}4\left(L-M+2l-2x+1\right)
\times \nonumber \\ & &\left\{{L+l-x+1 \choose M-k} {L+k\choose x-l+k}-{x+M-l\choose M-k}{L+k \choose x+M-l-1}\right\}\;\;\;\;,
\end{eqnarray}
which after some manipulation turns out to be
\begin{eqnarray}
\label{c6}
Z&=&4 \sum_{k=0}^{l-1}{l-1 \choose k}\sum_{M=0}^{L+2l}\left(M-k+1\right)\times \nonumber \\ & &
\left\{\sum_{x=0}^{{L-M\over 2}+l}{L+l-x+1 \choose M-k+1} {L+k\choose x-l+k}+
\sum_{x=0}^{{L-M\over 2}+l}{x+M-l\choose M-k+1}{L+k \choose x+M-l-1}\right\}\nonumber \\& &
-4 \sum_{k=0}^{l-1}{l-1 \choose k}\sum_{M=0}^{L+2l}\left(L+k\right)\times \nonumber \\ & &
\left\{\sum_{x=0}^{{L-M\over 2}+l}{L+l-x+1 \choose M-k} {L+k-1\choose x-l+k-1}+
\sum_{x=0}^{{L-M\over 2}+l}{L+k-1\choose x+M-l-1}{x+M-l \choose M-k}\right\}\;\;\;\;.
\end{eqnarray}
Substituting $x\to L-x-M+2l+1$ in the second and fourth terms of the equation above we get
\begin{eqnarray}
\label{c7}
Z&=&4 \sum_{k=0}^{l-1}{l-1 \choose k}\sum_{M=0}^{L+2l}\left(M-k+1\right)
\sum_{x=0}^{L-M+2l+1}{L-x+l+1 \choose M-k+1} {L+k\choose x-l+k}\nonumber \\& &
-4 \sum_{k=0}^{l-1}{l-1 \choose k}\sum_{M=0}^{L+2l}\left(L+k\right)
\sum_{x=0}^{L-M+2l+1}{L-x+l+1 \choose M-k} {L+k-1\choose x-l+k-1}
\end{eqnarray}
We now find
\begin{eqnarray}
\label{c8}
Z&=&-4 \sum_{k=0}^{l-1}{l-1 \choose k}\sum_{M=0}^{L+2l}{L+k \choose M-k}\frac{d}{dP}
\left[\sum_{x=0}^{L-M+2k}{L-M+2k \choose x-l+k}P^{x-l-L-1}\right]_{P=1}\nonumber \\& &
+4 \sum_{k=0}^{l-1}{l-1 \choose k}\sum_{M=0}^{L+2l}{L+k \choose M-k}\frac{d}{dP}
\left[\sum_{x=0}^{L-M+2k}{L-M+2k \choose x-l+k-1}P^{x-L-l-1}\right]_{P=1}
\end{eqnarray}
and using the binomial formula we obtain
\begin{equation}
\label{c9}
Z=4\frac{d}{dP}\left[\sum_{k=0}^{l-1}{l-1 \choose k}(1-P^{-1})P^{-k-L}\sum_{M=0}^{L+2l}{L+k \choose M-k}(1+P)^{L-M+2k}\right]_{P=1}\;\;\;\;.
\end{equation}
Above the sum over $M$ can be performed giving
\begin{equation}
\label{c10}
Z=4\frac{d}{dP}\left[(1-P^{-1})\sum_{k=0}^{l-1}{l-1 \choose k}P^{-k-L}(2+P)^{L+k}\right]_{P=1}.
\end{equation}
Now the sum over $k$ can be evaluated yielding
\begin{equation}
\label{c11}
4\frac{d}{dP}\left[(1-P^{-1})P^{-L-l+1}(2+P)^{L}(2+2P)^{l-1}\right]_{P=1}\;\;\;\;,
\end{equation}
which finally leads to
\begin{equation}
\label{c12}
Z=3^{L}4^{l}
.   \end{equation}
Thus we have shown that the number of eigenvectors of the
$t-J$ hamiltonian with
$l$ impurities is $3^{L}4^{l}$, which is precisely the number of states in the
Hilbert space of a chain of length $L$ with $l$ impurities. We point out that
this is the first time that completeness of the Bethe vectors
has been demonstrated for an impurity
 electronic model.

~~\\
~~\\
\centerline{{\bf Acknowledgements}}
~~\\

JL thanks the Funda\c{c}\~{a}o de Amparo a Pesquisa do Estado do Rio Grande do
Sul and Australian Research Council for financial support. He also thanks the
Instituto de F\'{\i}sica da UFRGS for their kind hospitality.
AF and APT thank CNPq-Conselho Nacional de Desenvolvimento Cient\'{\i}fico e
Tecnol\'ogico for financial support.


\end{document}